# The Evolution of Binaries Embedded Within Common Envelopes

Alejandra Rosselli-Calderon[1], Ricardo Yarza[1,2,6,7], Ariadna Murguia-Berthier[3,8], Valeriia Rohoza[1,3],  
Rosa Wallace Everson[1,9], Andrea Antoni[4], Morgan MacLeod[5], and Enrico Ramirez-Ruiz[1]  
[1] Department of Astronomy and Astrophysics, University of California, Santa Cruz, CA 95064, USA  
[2] Texas Advanced Computing Center, University of Texas, Austin, TX 78759, USA  
[3] Center for Interdisciplinary Exploration & Research in Astrophysics (CIERA), Evanston, IL 60202, USA  
[4] Astronomy Department and Theoretical Astrophysics Center, University of California, Berkeley, CA 94720, USA  
[5] Institute for Theory & Computation, Center for Astrophysics, Harvard & Smithsonian, Cambridge, MA, USA  
Received 2024 April 25; revised 2024 October 7; accepted 2024 October 7; published 2024 November 28

## Abstract

Triple stellar systems allow us to study stellar processes that cannot be attained in binary stars. The evolutionary phases in which the stellar members undergo mass exchanges can alter the hierarchical layout of these systems. Yet, the lack of a self-consistent treatment of common-envelope (CE) in triple-star systems hinders the comprehensive understanding of their long-term fate. This paper examines the conditions predicted around binaries embedded within CEs using local 3D hydrodynamical simulations. We explore varying the initial binary separation, the flow Mach number, and the background stellar density gradients as informed by a wide array of CE conditions, including those invoked to explain the formation of the triple system hosting PSR J0337+1715. We find that the stellar density gradient governs the gaseous drag force, which determines the final configuration of the embedded binary. We observe a comparable net drag force on the center of mass but an overall reduction in the accretion rate of the binary compared to the single-object case. We find that, for most CE conditions, and in contrast to the uniform background density case, the binary orbital separation increases with time, softening the binary and preventing it from subsequently merging. We conclude that binaries spiraling within CEs become more vulnerable to disruption by tidal interactions. This can have profound implications on the final outcomes of triple-star systems.

*Unified Astronomy Thesaurus concepts:* Common envelope evolution (2154); Common envelope binary stars (2156); Trinary stars (1714); Multiple stars (1081); Hydrodynamical simulations (767); Astronomical simulations (1857)

## 1. Introduction

Field stars are commonly formed in pairs, and many of these binaries are part of triples or even higher-order arrangements (I. A. Bonnell et al. 2003; M. Moe & R. Di Stefano 2017; H.-C. Hwang 2022). Even though the basics of single-stellar evolution and binary evolution have largely been established, many aspects of triple-star evolution remain mysterious (S. Toonen et al. 2016).

Triples are of general interest as progenitors of compact binaries (T. A. Thompson 2011; F. Antonini et al. 2016). This class includes hot subluminous stars (H. P. Preece et al. 2022), ultra-compact X-ray binaries (S. Naoz et al. 2016), double white dwarfs (A. S. Rajamuthukumar et al. 2023), and double neutron stars (A. S. Hamers & T. A. Thompson 2019). Compact binaries are central to understanding type Ia supernovae and short gamma-ray burst explosions (T. A. Thompson 2011; A. S. Rajamuthukumar et al. 2023). Of particular interest are triple systems that serve as potential progenitors for LIGO binaries (F. Antonini et al. 2016; A. Vigna-Gómez et al. 2021), which are spectacular probes of the extreme physics at high energy and high density, being strong gravitational-wave sources. Triples have been found to harbor millisecond pulsars (S. M. Ransom et al. 2014), and as such are ideal test beds of general relativity and alternative theories of gravity (A. M. Archibald et al. 2018).

Observed stable triple systems are mostly found in a hierarchical arrangement. These systems are composed of two closely orbiting stars in an inner binary, hereafter referred to as the binary, and an outer star in a wider orbit about the center of mass of the triple. For the cases explored in this work, the outermost star is the more massive of the three, and hereafter will be called the primary. A critical juncture in the life of a triple star is the period just after mass transfer commences in the system (S. Toonen et al. 2016; E. Michaely & H. B. Perets 2019; T. A. F. Comerford & R. G. Izzard 2020; H. Glanz & H. B. Perets 2021a; N. Soker & E. Bear 2021). The system either merges or may survive to become an interacting triple. When mass transfer is unstable, the binary is believed to shrink. During this common envelope (CE) phase, the envelope of the donor star engulfs both stars, causing the binary to spiral inward (B. Paczynski 1976). This tightening occurs due to the influence of drag from the surrounding envelope material. This gravitational drag force deposits orbital kinetic energy into the envelope and is expected to influence the orbit of the infalling binary, whose orbital evolution remains poorly understood. If the embedded binary spirals deep enough within the envelope of the evolving star, it can deposit sufficient energy to unbind the entire hydrogen envelope (T. A. F. Comerford & R. G. Izzard 2020). The outcome of this envelope ejection can lead to the core of the evolving star being placed in a

---

substantially tightened triple system—or binary system, if one of the binary members is tidally ejected. The millisecond pulsar triple system PSR J0337+1715 has been theorized to be one of such systems where a CE phase could have played a significant role in its evolution and final outcome (T. M. Tauris & E. P. J. van den Heuvel 2014; E. Sabach & N. Soker 2015; F. Lagos et al. 2023). This system is used here to guide our simulation work.

The detailed understanding of binaries within a CE remains largely unexplored (S. Toonen et al. 2016; T. A. F. Comerford & R. G. Izzard 2020), primarily because the interaction is governed by intertwined physical processes on a large range of scales (H. Glanz & H. B. Perets 2021a, 2021b). The governing timescales range from the stellar evolutionary timescale that establishes the initial conditions of a binary interaction to the dynamical timescale of flow near an embedded objects surface. These challenges imply that global simulations can not effectively capture the full range of physical processes that determine the evolution of a binary system during a CE interaction.

A specific concern is that the full spatial range of the embedded binary for many orbital timescales indicates that the calculations are either highly computationally expensive or are achieved with exceedingly low numerical resolution. In this study, we adopt the complementary method of simulating a well-defined but idealized scenario. We study the flow past a binary system in the context of a wind tunnel numerical setup, detailed in Section 4 (M. MacLeod & E. Ramirez-Ruiz 2015a; M. MacLeod et al. 2017; A. Antoni et al. 2019). To determine the conditions of the flow, we consider stellar structures (and gas adiabatic exponents) appropriate to a wide range of regimes of CE interactions in hierarchical triples. By restricting the scope of the problem, this idealized approach allows us to, for the first time, study the orbital evolution of binaries embedded in stellar envelopes.

This paper is organized as follows. In Section 2, we present the case study of the remarkable triple system hosting PSR J0337+1715, which informs the parameter space we explore. In Section 3, we review the key descriptive parameters for CE flows and discuss the relevant range of values with a particular emphasis on the assembly of PSR J0337+1715. These flow properties are used to inform the setup of our numerical experiments. Section 4 describes our numerical method while Section 5 examines the results of a set of numerical experiments. Section 6 summarizes the implications that these idealized results have for our understanding of the evolution of embedded binary systems undergoing typical CE interaction.

## 2. Case Study: PSR J0337+1715

As an example of the typical flow properties expected for binaries embedded in CEs, we discuss here one of the commonly invoked formation scenarios for the remarkable triple-star system that harbors the millisecond pulsar PSR J0337+1715 (T. M. Tauris & E. P. J. van den Heuvel 2014; E. Sabach & N. Soker 2015). This system is comprised of a $1.438 M_\odot$ millisecond pulsar with a spin period of 2.73 ms with two white dwarfs companions (S. M. Ransom et al. 2014; A. M. Archibald et al. 2018). The white dwarf member of the inner binary has a mass of $0.1975 M_\odot$ and a semimajor axis of $0.83 R_\odot$, while the distant white dwarf orbiting the center of mass of the inner binary has a mass of $0.41 M_\odot$ and a semimajor axis of $51.28 R_\odot$.

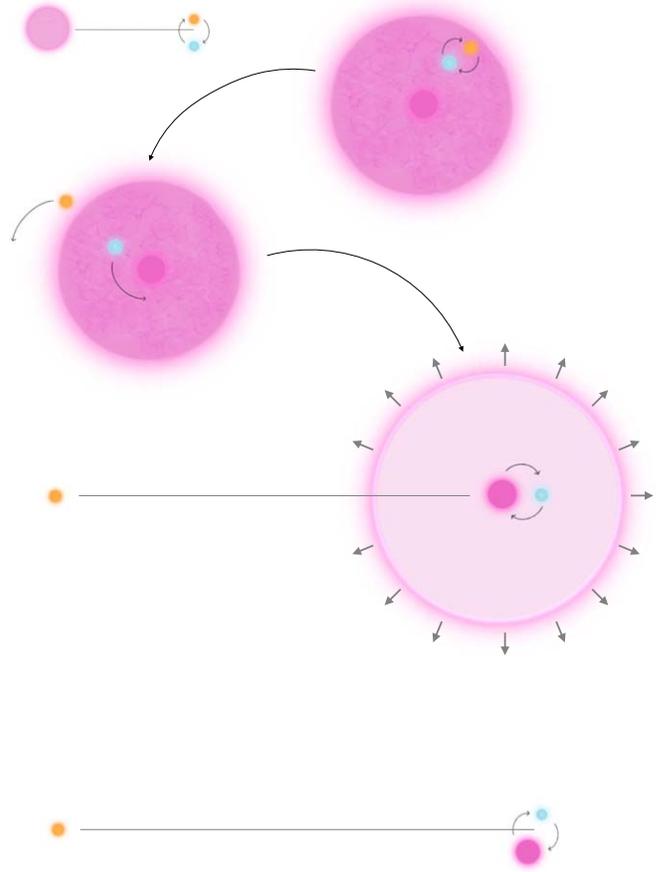

**Figure 1.** Diagram illustrating the proposed post-CE formation scenario for PSR J0337+1715 (T. M. Tauris & E. P. J. van den Heuvel 2014; E. Sabach & N. Soker 2015). The initial setup consists of three main-sequence stars: a primary (pink) with $M \approx 10 M_\odot$, a secondary (blue), and a tertiary (orange) with $m_1 \approx m_2 \approx 1 M_\odot$, where $M > m_1 + m_2$. As the primary evolves, it will expand and the system will undergo a CE. The center of mass of the binary will sink toward the core of the primary until it reaches its tidal radius. The tertiary is ejected from the CE but remains bound to the system, while the secondary sinks and unbinds the envelope. This results in a hierarchical triple system in which the inner binary is composed of the primary's core and the secondary.

In some triple systems, the outer star has no influence on the evolution of the inner binary, such that the evolution of the inner binary and the outer tertiary can be explained independently. However, in this case, it is reckoned that interactions between the three stars took place early in the evolution of the system (T. M. Tauris & E. P. J. van den Heuvel 2014), and as such have all been influenced in their evolution by their companions. As the outer stellar progenitor of PSR J0337+1715's expanded, it has been suggested that the inner Sun-like pair was completely engulfed, causing the inner binary to spiral inward (E. Sabach & N. Soker 2015).

Based on the scenario described in E. Sabach & N. Soker (2015) and T. M. Tauris & E. P. J. van den Heuvel (2014), we present a tentative history of the system containing PSR J0337 +1715. In Figure 1, we show a schematic montage of successive interactions that are believed to be central to the formation of PSR J0337+1715. The associated frames represent educated guesses of their geometrical arrangements. Let us consider these frames in turn, working from the first binary interaction to the post-CE formation configuration. The proposed formation scenario for PSR J0337+1715 begins with a primary massive main-sequence star (pink star in Figure 1) orbiting a tight binary composed of two Sun-like stars. As the outer massive star progenitor of PSR J0337+1715's expands, it is speculated that the inner Sun-like pair will be completely engulfed. As the binary spirals inside of the primary's envelope, it is predicted to be tidally disrupted. This will result in the ejection of one of the Sun-like companions (orange star in Figure 1). After the binary is tidally separated, the surviving companion (light blue star in Figure 1), continues to spiral in toward the primary's He core, leading to the successful ejection of the H envelope. The post-CE configuration of the triple system is predicted to host a close binary, hosting the He core and the surviving Sun-like companion, along with the outer tertiary companion, which is expected to remain gravitationally bound. In what follows, we study the common envelope flow conditions that the binary would have encountered during the CE phase as illustrated here.

### 2.1. CE Evolution Within the Triple System PSR J0337+1715

The setup of the problem is as follows. We have a primary star with total mass $M$ and radius $R$ that engulfs its two less massive companions $m_1$ and $m_2$, arranged in a close binary. We study the case where $m_1 = m_2$ and define the total mass of the binary system as $m = m_1 + m_2$. The separation between $m_1$ and $m_2$ is given by $a_{b,0}$. The distance between the center of mass of the binary and the center of mass of the primary is defined as $a$, where $a \lesssim R$. For the cases relevant to this study, $a_{b,0} < R$, which ensures that both members of the binary are embedded within the primary's stellar envelope. This is expected to be the case for most triples, which are observed to have hierarchical structures (S. Toonen et al. 2016). We define the mass ratio between the primary and the embedded binary system as $q = m/M$.

The upper panel of Figure 2 shows the evolutionary history, as envisioned by E. Sabach & N. Soker (2015) and T. M. Tauris & E. P. J. van den Heuvel (2014), of the primary stellar progenitor of PSR J0337+1715. We use the stellar evolution code MESA (version r24.03.01; B. Paxton et al. 2011, 2013, 2015, 2018, 2019; A. S. Jermyn et al. 2023) to model the primary star, with an initial mass of $M = 10 M_\odot$ and solar metallicity, from zero-age main sequence to the asymptotic giant branch. In the first stage of the evolution of the system, the primary's envelope will embed the inner binary. For simplicity, we assume the binary hosts two Sun-like stars, such that $m_1 = m_2 = 1 M_\odot$ and $q = 0.2$.

As the binary spirals inside of the primary's envelope, it is expected to be tidally disrupted, resulting in the ejection of one of the Sun-like companions. The tidal disruption radius of the binary can be written as

$$r_{\tau,b} = a_b \left( \frac{M_{enc}}{2m} \right)^{1/3}, \quad (1)$$

where $M_{enc}$ is the enclosed mass within the orbit of the center of mass of the binary and $a_b$ is the separation of the binary at the time of disruption, which might be modified as the binary sinks (Section 5.3). We find this disruption to occur at a radius of $\approx 70 R_\odot$ from the core of the primary for a binary separation $a_{b,0} = 60 R_\odot$ (orange curve in the lower panel of Figure 2). After the binary breakup, the non-ejected companion, which is still embedded within the CE, continues to spiral in toward the primary's core (pink curve in the lower panel of Figure 2). It is then up to this companion to aid with the ejection of the envelope, which is expected to take place within the blue shaded region in the lower panel of Figure 2.

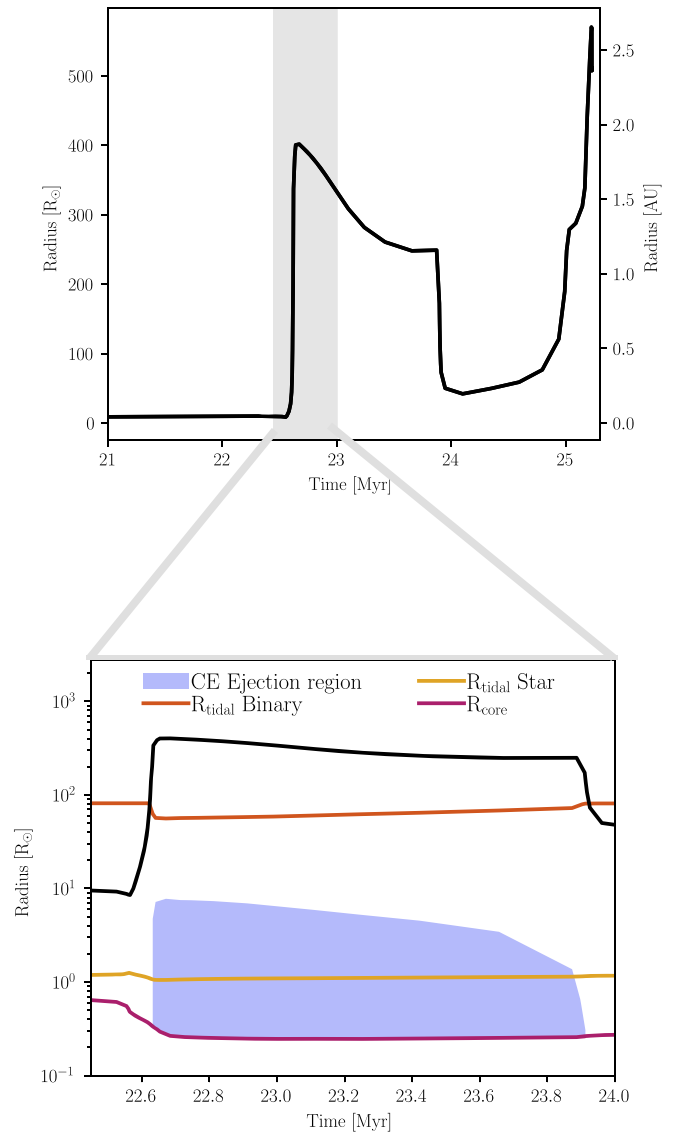

**Figure 2.** Profiles of primary-star stellar structure in a triple system relevant to CE interaction. Top panel: the radial evolution of the $10 M_\odot$ progenitor star of PSR J0337+1715 is shown as a function of time. The outer binary, assumed here to be composed of two $1 M_\odot$ Sun-like stars, is expected to be embedded within the envelope of the primary star, provided that it resides at separations below $400 R_\odot$. The vertical axis shows the radial extent of the primary in $R_\odot$ units (left) and astronomical units (right). Bottom panel: the internal structure of the primary as calculated by MESA. The radial extent of the primary is shown in black. As the binary spirals, it is expected to be tidally disrupted at a radius given by the orange curve, while the yellow curve represents the tidal disruption radius of a single $1 M_\odot$ Sun-like star. The shaded blue region represents the range of radii in which CE ejection is possible due to a single Sun-like star, since the binary will have already been tidally disrupted at the radius represented in orange. The pink curve represents the outer edge of the primary's He core.

The shaded blue region in Figure 2 is calculated using the standard $\alpha$-formalism (E. P. J. van den Heuvel 1976; R. F. Webbink 1984), where one compares the orbital energy deposited by the binary onto the envelope of the primary ($\Delta E_{orb}$) to the binding energy of the envelope ($E_{bind}$). We define the deposited orbital energy as

$$\Delta E_{\rm orb} = \frac{GMm}{2a_i} - \frac{GM_{\rm enc}m}{2a_f}, \quad (2)$$

in which $a_i$ is the initial separation and $a_f$ is the final separation between the center of mass of the binary and the center of mass of the primary's core. We define the envelope's binding energy as

$$E_{\rm bind} = \int_{M_{\rm core}}^{M} -\frac{GM(a)}{a} dm, \quad (3)$$

with $M_{\rm core}$ being the mass of the He core as calculated from the MESA profiles and $M(a)$ being the enclosed mass at $a$. In order to define a complete envelope ejection, we find the region where $\alpha \Delta E_{\rm orb} \geqslant E_{\rm bind}$, and use a value of $\alpha = 1.0$. For additional details on the choice of $\alpha$ and the implementation of the $\alpha$ formalism when using MESA profiles, the reader is referred to J. A. P. Law-Smith et al. (2020), S. Wu et al. (2020), R. W. Everson et al. (2024, Ch. 3), and R. W. Everson et al. (2024).

The dynamics and fate of binaries embedded in a CE have not been studied in great detail, as it is an inherently complicated problem where the dynamics of the binary's orbit and the hydrodynamics of the flow need to be taken into account self-consistently. The case of PSR J0337+1715 offers a conceptual template for defining the flow properties that may be typically encountered by embedded binaries in hierarchical triples. In the following section, we calculate these flow properties during the dynamical inspiral phase, which is expected to terminate when the envelope is successfully ejected (shaded blue region in Figure 2). The tidal disruption of the binary results in the ejection of one of the binary members at an outer radius. The radius range in which the remaining member of the binary may cause CE ejection is well outside the tidal disruption radius of that member, due to the primary's core. This is promising, given that the formation of PSR J0337+1715 necessitates both envelope ejection and the survival of all triple members (Figure 1). It is important to note that these ranges are typical of those expected for the vast majority of CE encounters (R. W. Everson et al. 2020) and will be used to guide our numerical setup, which is presented in Section 4.

## 3. Flow Conditions

When modeling the flow around the embedded binary, we parameterize the properties of the gas using the formalism developed by M. MacLeod & E. Ramirez-Ruiz (2015a, 2015b) and M. MacLeod et al. (2017). In this section, we explore the general conditions of the flow and provide the reader with flow parameters relevant to the assembly history of PSR J0337+1715.

### 3.1. Flow Properties During CE

The flow can be contextualized with the characteristic scales of a Bondi–Hoyle (BH) accretion (F. Hoyle & R. A. Lyttleton 1939; H. Bondi & F. Hoyle 1944). The accretion capture radius $R_a$ of the binary can be written as

$$R_a = \frac{2Gm}{v_\infty^2} \quad (4)$$

for supersonic flows, where $m$ is the total mass of the binary and $v_\infty$ is the relative velocity of the center of mass of the binary system and its surrounding medium. We follow the prescription detailed in Section 2 of A. Antoni et al. (2019) to derive the following relations between the flow parameters. The accretion capture radius will inform the rate of accretion of the two stars. The Mach number of the flow is $\mathcal{M}_\infty = v_\infty/c_{s,\infty} \gtrsim 1$, where $c_{s,\infty}$ is the sound speed of the gas. These imply an accretion rate onto the embedded binary,

$$\dot{m}_{\rm BH} = \pi R_a^2 \rho_\infty v_\infty \left( \frac{\mathcal{M}_\infty^2}{1 + \mathcal{M}_\infty^2} \right)^{3/2}, \quad (5)$$

where $\rho_\infty$ describes the characteristic background density. The associated orbital energy decay rate is then

$$\dot{E}_{\rm BH} = \frac{\pi}{2} R_a^2 \rho_\infty v_\infty^3 \left( \frac{\mathcal{M}_\infty^2}{1 + \mathcal{M}_\infty^2} \right)^{1/2}, \quad (6)$$

which leads to a characteristic stopping timescale of

$$\tau_{\rm stop,BH} = \frac{\frac{1}{2}mv_\infty^2}{\dot{E}_{\rm BH}} = \frac{v_\infty^3}{4\pi G^2 m \rho_\infty} \left( \frac{\mathcal{M}_\infty^2}{1 + \mathcal{M}_\infty^2} \right)^{1/2}. \quad (7)$$

The orbital energy decay traces the slowing down of the center of mass of the binary. The characteristic orbital velocity of the binary system around the primary star is

$$v_k = \sqrt{\frac{G(M + m)}{a}}. \quad (8)$$

In general, $v_\infty \approx v_k$, such that $\tau_{\rm stop,BH} \approx (E_{\rm orb}/\dot{E}_{\rm BH})$ confers the radial sinking timescale of the binary in the stellar envelope, where we have assumed that the mean stellar density enclosed by the orbit is $\approx \rho_\infty$. The motion of the embedded binary might be, however, desynchronized from the primary's gaseous envelope, and the relative velocity can be written as $v_\infty = f_k v_k$, where $f_k$ is the fraction of Keplerian velocity that describes the relative motion between the center of mass of the binary and the gas. $R_a$ can then be expressed as

$$\frac{R_a}{a} = \frac{2}{f_k^2} \frac{m}{(M + m)} = \frac{2}{f_k^2} \frac{1}{(1 + q^{-1})}. \quad (9)$$

In the case of $f_k = 1$ and $M \gg m$, then $(R_a/a) \approx 2q$. It is important to note that when the enclosed primary-star mass, $M(a)$, is considerably less than $M$, one needs to use $q_{\rm enc} = m/M(a)$ in place of $q$ in Equation (9).

Another characteristic length scale that plays a key role in defining the CE interaction is the density scale height (M. MacLeod & E. Ramirez-Ruiz 2015a, 2015b), which depends on the structure of the primary star and is given by

$$H_\rho = -\rho \frac{dr}{d\rho}. \quad (10)$$

For the case of the embedded binary, we measure $H_\rho$ at the center of mass of the system. For simplicity, we define a dimensionless parameter $\varepsilon_\rho$ to describe the number of density scale heights encapsulated by the accretion radius:

$$\varepsilon_\rho \equiv \frac{R_a}{H_\rho}. \quad (11)$$

For the case of a binary moving through a uniform density medium $\varepsilon_\rho = 0$, while $\varepsilon_\rho \gtrsim 1$ describes relatively steep stellar density gradients (M. MacLeod & E. Ramirez-Ruiz 2015a; A. Antoni et al. 2019; R. W. Everson et al. 2020). The vertical gradient, whose strength is determined by $\varepsilon_\rho$, provides the flow net angular momentum relative to the accreting object (A. Murguia-Berthier et al. 2017). Thus, we expect that even modest gradients could have sizable impacts on the evolution of the orbit of the embedded binary.

The envelope and the binary will both be subjected to the gravitational force of the primary star, which allows one to connect the properties of the CE flow on the basis of the hydrostatic equilibrium conditions of the stellar envelope (M. MacLeod et al. 2017):

$$\mathcal{M}^2 = \varepsilon_\rho \frac{(1+q)^2}{2q} f_k^4 \left( \frac{\Gamma_s}{\gamma} \right), \quad (12)$$

where $\Gamma_s$ is the polytropic index of the stellar profile and $\gamma$ is the gas adiabatic index. This relation is particularly useful because it limits the relevant parameter space that one needs to explore numerically to effectively characterize the CE flows. In the stellar envelopes in hydrostatic equilibrium for massive stars in giant phases, the ratio of $\Gamma_s$ to $\gamma$ will be $\approx 1$. For a fixed $f_k$, the Mach number $\mathcal{M}$ depends solely on $\varepsilon_\rho$ and $q$, allowing us to vary these two parameters and draw aggregate conclusions (R. W. Everson et al. 2020).

### 3.2. Flow Conditions Expected During CE in Triple System PSR J0337+1715

Using the triple system PSR J0337+1715 as a base case scenario, the specific values for the flow parameters for a binary system embedded in an evolved massive star are discussed. Based on MESA models, here we calculate $\varepsilon_\rho$ as a function of the radial distance from the core of the evolving primary star described in Section 2. When the primary enters the giant phase, the envelope will engulf the orbiting binary. This is the evolutionary phase that is key to the assembly of PSR J0337 +1715 (T. M. Tauris & E. P. J. van den Heuvel 2014; E. Sabach & N. Soker 2015). We calculate that the primary will expand up to a radius of about $400 R_\odot$ during the asymptotic giant branch (Figure 2).

As the primary star begins to transfer mass at the Roche limit, we expect a phase of runaway decay leading up to the plunge of the binary within the envelope of the donor. According to M. MacLeod & A. Loeb (2020), the plunge of the binary is predicted to initiate at $a \approx 300 R_\odot$, while the runaway decay will take place bounded by $\Delta a \approx (400–300) R_\odot \approx 100 R_\odot$. During this period, the binary is fully embedded into the envelope, given that $a_{b,0} < R_a \lesssim \Delta a$.

In Figure 3, we illustrate the relevant flow conditions for this representative model, bounded by the limits of onset and binary disruption radii. We anticipate the tidal disruption of the binary ($a_{b,0} = 60 R_\odot$) to take place before the center of mass reaches the innermost $70 R_\odot$ of the primary star (orange line in Figure 3). This will likely result, as envisioned by E. Sabach & N. Soker (2015), in the ejection of one of the Sun-like companions. After the binary breaks up, the remaining bound companion will continue to spiral in toward the primary's core. During this phase, we expect enough energy to be injected by the surviving companion to unbind the surrounding envelope before its reaches the tidal radius (yellow line in Figure 3). In this range of radii, we determine the value of the density gradient parameter, $\varepsilon_\rho$, to be between 1.0 and 2.1, and the value of the Mach number, $\mathcal{M}$, to be between 1.47 and 2.74. These value ranges of $\varepsilon_\rho$ and $\mathcal{M}$ are indeed representative of most CE flow conditions (R. W. Everson et al. 2020).

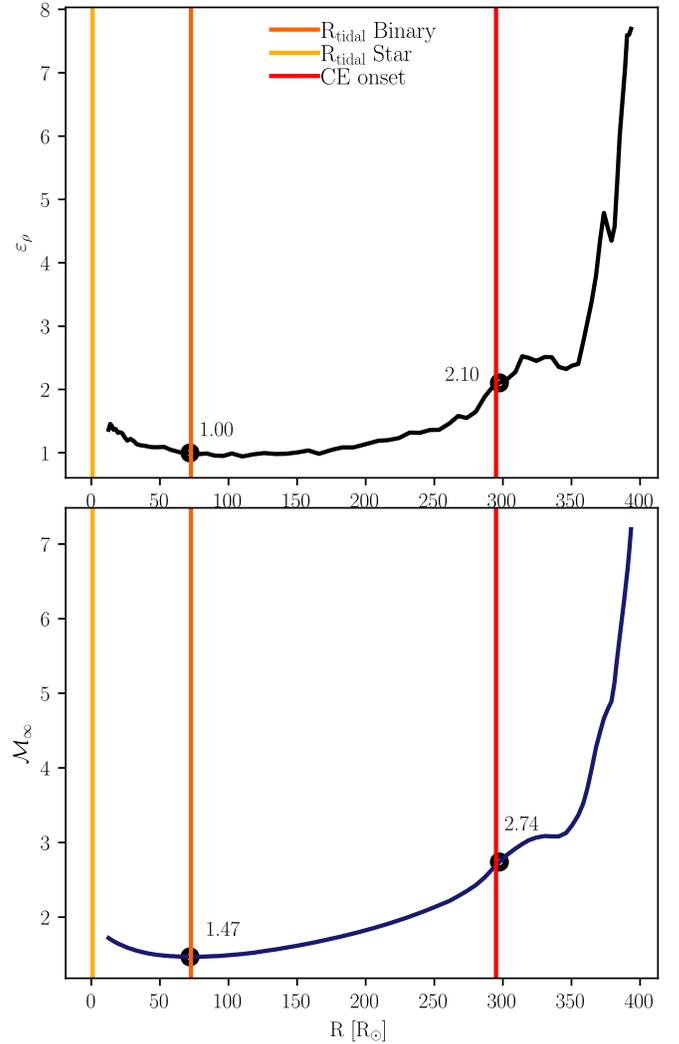

**Figure 3.** Common envelope flow conditions during the assembly of PSR J0337+1715. $\varepsilon_\rho$ (top panel) and $\mathcal{M}_\infty$ (bottom panel) are each shown as a function of the radius from the center of mass of the primary during the asymptotic giant branch. The yellow line represents the radius at which a 1 $M_\odot$ Sun-like star will become tidally disrupted, while the orange line shows the tidal disruption radius for the embedded binary ($a_{b,0} = 60 R_\odot$). The red line shows the CE onset boundary as defined in M. MacLeod & A. Loeb (2020), implying that most of the material above this line will be lost before CE sets in.

During the dynamical sinking of the binary into the envelope of the primary, we have to consider under which conditions the local simulations presented in this paper are able to effectively capture the global dynamics. For this assumption to be defensible, the binary must encounter undisturbed envelope material as it sinks toward the core. As a simple approximation to check whether this is the case, we follow M. MacLeod & E. Ramirez-Ruiz (2015a) and compare the stopping timescale with the orbital period of the binary's center of mass around the primary's core. We use the commonly used parameter $\beta_{CE} = (E_{orb}/\dot{E}_{BH})/P_{orb}$ as defined in M. Livio & N. Soker (1988). When $\beta_{CE} \lesssim 1$, local effects dominate, since the binary sinks in faster than it orbits around the primary's core. This implies that the embedded binary will not encounter the wake of its shock from the previous orbit, thus justifying the use of local simulations (M. MacLeod & E. Ramirez-Ruiz 2015a). For our case study, $\beta_{CE} \lesssim 0.6$ for all separations. The accretion radius of the binary at a distance $a = 300 R_\odot$, when CE onset is expected to occur, is $\sim 100 R_\odot$, such that $a_{b,0}/R_a = 0.6$. As the binary continues to sink, $\beta_{CE}$ systematically decreases.

The system's post-CE configuration is expected to leave the core of the primary and secondary in a close binary and the tertiary companion further out but still gravitationally bound. While the system will further evolve and additional interactions are expected to occur (T. M. Tauris & E. P. J. van den Heuvel 2014; E. Sabach & N. Soker 2015), we note that one of the major uncertainties in the modeling comes from the lack of understanding of the evolution of the inner Sun-like pair and the poorly known orbital evolution of the binary during the CE phase.

## 4. Numerical Approach

The wind tunnel is a three-dimensional, Cartesian geometry hydrodynamic simulation setup that includes the gravity of the embedded binary and the primary evolving star, which allows us to study the coefficients of drag and accretion experienced by the embedded binary. Accretion and drag lead to a transformation of both the center of mass of the binary and its orbit during a CE phase.

### 4.1. Wind Tunnel Conditions

As the upstream conditions become inhomogeneous, the symmetry that defines Bondi–Hoyle accretion is quickly broken. Instead, we observe an asymmetric flow, in which the momenta of incoming fluid no longer cancel in the wake (M. MacLeod & E. Ramirez-Ruiz 2015a). These flow properties can alter the rates of drag of the sinking of the center of mass of the system and the evolution of the orbit of the embedded binary when compared with the Hoyle–Lyttleton case (A. Antoni et al. 2019).

We use the FLASH hydrodynamics code (B. Fryxell et al. 2000) to numerically solve the fluid equations. We implement FLASHs directionally split Piecewise Parabolic Method Riemann solver (P. Colella & P. R. Woodward 1984). We use sinks (C. Federrath et al. 2010) to define the stars as point masses that accrete material from their environment and respond to the forces induced by the gas. We place two orbiting sinks in the gas medium and allow them to move freely, unconstrained to a fixed center of mass location (A. Antoni et al. 2019). The numerical formalism is described in detail in M. MacLeod et al. (2017), but for context we will discuss some key aspects of the problem setup.

A wind, depicting the stellar envelope of the primary star, is injected in the $+x$ direction from the $-x$ boundary. This wind has a vertical gradient of density and pressure, whose structure is calculated using the 1D stellar profile under the assumption of hydrostatic equilibrium. To be exact, we introduce a vertical $-y$ acceleration given by the gravitational influence of the enclosed mass of the primary star, which remains uniform in the $z$ direction throughout the simulations. The conditions of the flow are parameterized by $\varepsilon_\rho$, $q$, and the pressure and density at $y = 0$ determined by $f_k$, $\Gamma_s$ and $\gamma$ (Equation (12)). The values at $y = 0$ are then used to inform the vertical structure of the flow $\pm y$, which is calculated assuming hydrostatic equilibrium.

In this setup, we model the embedded binary ($m_1 = m_2$) as two sink point particles of radius $R_s = 0.05 R_a$, whose center of mass at $t = 0$ is at rest ($V_{CoM} = 0$) and placed at the origin of the grid ($R_{CoM} = 0$). The binary is originally placed in a circular orbit at a distance $a_{b,0}$ in the $x$–$y$ plane with the angular momentum vector pointing in the $+z$ direction. As the system evolves in time, $V_{CoM}$ and $R_{CoM}$ change as they accrete material and interact gravitationally with each other and the background gas. We perform a resolution study by varying $R_s$ and obtain numerical convergence at scales consistent with those found in A. Antoni et al. (2019), which we use throughout our study ($R_s = 0.05 R_a$). For additional details on the resolution study and the binary implementation, the reader is referred to A. Antoni et al. (2019).

The simulations are performed in dimensionless units, where the velocity of the incoming wind at $y = 0$ is set to $v_\infty = 1$. By making this choice, we are fixing the value of the mass of the embedded binary to be $m = (2G)^{-1}$. This naturally yields a dimensionless accretion radius $R_a = 2Gm/v_\infty^2 = 1$. The density of the stellar envelope at a radial distance $a$ ($y = 0$) from the center of mass of the primary star is characterized by $\rho_\infty = 1$. In code units, the characteristic timescale is $R_a/v_\infty = 1$ and the characteristic length scale is $R_a$. Previous work has studied the effects of domain size and spatial resolution on the robustness of the numerical results for single objects embedded in CEs (M. MacLeod et al. 2017) and binaries moving supersonically relative to a uniform medium (A. Antoni et al. 2019). Based on these studies and our own resolution study, we adopt the following bounds for our simulation runs. The computational domain is $(-3, 3)R_a \times (-3, 3)R_a \times (-3, 3)R_a$. The maximum and minimum refinement levels are set to 6 and 2, respectively. Therefore, the maximum cell size is $R_a/16$ and the minimum is $R_a/256$.

The parameters that we vary in the simulations are selected to investigate the evolution of embedded CE binaries with a range of orbital periods and experiencing a wide range of stellar density gradients. As such, we change $\varepsilon_\rho$ and $a_{b,0}$ between 0 and 2.0, and 0.16 to $1.0 R_a$, respectively. Throughout the simulations, we assume $\Gamma_s = \gamma = 5/3$ (R. W. Everson et al. 2020) and perform calculations with $q = 0.1$ and $q = 0.15$ (Section 2). By selecting $q$ and $\varepsilon_\rho$, the upstream Mach number ($\mathcal{M}_\infty$) can be computed employing Equation (12). The use of dimensionless flow parameters in our simulations allows for the results of this study to be broadly applicable to a wide range of embedded binaries (Appendix). This is because the properties of the flow throughout CE encounters are self-similar across a broad range of post-main-sequence primary masses, evolutionary stages, mass ratios, and metallicities (R. W. Everson et al. 2020).

### 4.2. Quantifying the Role of Drag and Mass Accretion

The evolution of the embedded binary depends sensitively on gas drag and mass accretion, which slows the center of mass and alters orbital spiraling. When the gas approaches the boundaries of a sink, it gets accreted. The accretion step is performed by integrating a quantity, either mass or momentum, over all cells within the volume of the sink and then adding these summed values to the particle's properties (A. Antoni et al. 2019). To compute the rate of accretion, we then divided the accreted quantity by the time step. The accretion rate is saved and then the gas is deleted by resetting the density within the sink to $\rho_{\text{sink}} = 10^{-2}\rho_\infty$ as well as setting the velocity components of the gas to zero. The mass accretion rate is calculated as a volume integral over a sink $i$ such that

$$\dot{m}_i = \frac{1}{\Delta t} \int_{\text{sink}_i} (\rho - \rho_{\text{sink}}) \, dV. \tag{13}$$

The total accretion rate of the binary is then the sum of the individual accretion rates of the sinks, $\dot{m} = \dot{m}_1 + \dot{m}_2$.

The accretion for the linear momentum of the gas onto the sink is integrated in each Cartesian coordinate as

$$\dot{\boldsymbol{p}}_i = \frac{1}{\Delta t} \int_{\text{sink}_i} \boldsymbol{v}(\rho - \rho_{\text{sink}}) \, dV. \tag{14}$$

The accretion of linear momentum will exert a force on the sinks. We call these forces "momentum transport forces" and denote them as $\boldsymbol{F}_{\dot{p}_1}$ and $\boldsymbol{F}_{\dot{p}_2}$, acting on $m_1$ and $m_2$, respectively (A. Antoni et al. 2019). The net momentum transfer force on the center of mass of the binary will be defined as

$$\boldsymbol{F}_{\dot{p}} = \boldsymbol{F}_{\dot{p}_1} + \boldsymbol{F}_{\dot{p}_2}. \tag{15}$$

As the sinks evolve over time, they exert a gravitational force on the surrounding gas. These forces cause the gas to get redistributed. In response to this, the gas exerts a gravitational force on each of the members of the binary. The total force on each member can be found by summing over the contributions of each of the cells in the domain. Using $\boldsymbol{r}_i$ as the location of a particle $m_i$ in the domain and $\boldsymbol{r}'$ as the location of a particular gas cell, we can find the force on $m_i$ due to the cell as

$$d\boldsymbol{F}_{\text{DF}_i} = -\frac{Gm_i \rho(\boldsymbol{r}') dV}{|\boldsymbol{r}_i - \boldsymbol{r}'|^3}(\boldsymbol{r}_i - \boldsymbol{r}'). \tag{16}$$

We take the integral over a sphere of volume $V$ with center at the center of mass of the binary to find the total gas drag force on $m_i$:

$$\boldsymbol{F}_{\text{DF}_i} = -\int_V \frac{Gm_i \rho(\boldsymbol{r}') dV}{|\boldsymbol{r}_i - \boldsymbol{r}'|^3}(\boldsymbol{r}_i - \boldsymbol{r}'). \tag{17}$$

The net gas drag force on the center of mass is

$$\boldsymbol{F}_{\text{DF}} = \boldsymbol{F}_{\text{DF}_1} + \boldsymbol{F}_{\text{DF}_2}. \tag{18}$$

The position and corresponding velocity of each particle are updated to account for the accreted material. Before the gas is evolved further, each particle's motion is advanced using the active sink particle leapfrog integrator (A. Antoni et al. 2019). At each time step, we document the accretion rates and forces, in addition to the position, total mass, position, and velocity of each particle.

### 4.3. Flow Morphology

We illustrate here the morphology of the CE flow around the binary for the suite of simulations used in this study. The snapshots are taken at the orbital plane $(z = 0)$, and the colormap shows the density of the flow. The flow lines are drawn along velocity streamlines. Figure 4 plots nine different configurations of initial conditions. The three different columns from left to right show the change in initial semimajor axis: $a_{b,0} = 0.16$ (left), 0.42 (middle), and 1.0 (right). The three different rows show changes in $\varepsilon_\rho$: 0.5 (top), 1.0 (middle), and 2.0 (bottom). For the cases where the binary separation is the smallest, we see a clearly defined spiral shock structure. This is likely due to the bodies orbiting one another much faster than the flow crossing time.

As the value of $\varepsilon_\rho$ increases, the flow is observed to exhibit significantly variability imprinted at the smallest scales. This takes place as vortices shed in the wake cause some degree of breathing and instability of the location of the bow shock. Since this flow instability is not observed in the shallower-gradient cases (M. MacLeod et al. 2017), the resultant drag force is much less variable. What is more, for large $\varepsilon_\rho$, there is a decrease in concentration of gas around the sinks, which drastically decreases the torque pushing the binary inward (A. Antoni et al. 2019). The majority of the net torque pushing the binary outward is generated at larger scales, on the order of the shock standoff distance.

### 5. Numerical Results

Our study aims to address questions surrounding the morphology and structure of flows in the vicinity of a binary embedded within a CE, the relative rates of drag and accretion, and how those flows influence the orbital evolution of the embedded system. These results are compared for robustness with those from a single object embedded within a CE $(a_{b,0} = 0)$ and a binary moving supersonically relative to a uniform medium $(\varepsilon_\rho = 0)$, which have been extensively studied by M. MacLeod et al. (2017) and A. Antoni et al. (2019), respectively.

### 5.1. Binaries Embedded in CEs

We begin examining the contributions of the gas dynamical friction and the accretion of linear momentum by the sink boundaries to the net drag force. The effect on the binary is compared with the case in which the system is replaced by an effective star at the center of mass of the binary $(a_{b,0} = 0)$. For this set of simulations, the initial separation of the binary is set to $a_{b,0} = 0.42R_a$, the density gradient parameter is $\varepsilon_\rho = 0.3$, and the Mach number $\mathcal{M} \approx 1.0$. Figure 5 shows the net drag forces including contributions from dynamical friction (Equation (18)) and accretion of linear momentum (Equation (15)) projected onto the velocity of the center of mass vector. The solid lines in the left panel of Figure 5 show the net force on the center of mass of the binary, while the dotted line shows the net force on the effective single-mass object. The gravitational drag force, given by Equation (18), is integrated over four different volumes for each time step, following the procedure described in M. MacLeod et al. (2017). The integration volumes are selected to be spherical shells with their inner radii placed at the center of mass of the binary and varying outer radii between $0.78R_a$ and $3R_a$. The different colors in the left panel of Figure 5 correspond to different outer radii $(r_{\text{out}})$. As expected, the gas dynamical friction forces grows as $\propto \ln(r_{\text{out}})$.

Figure 5 shows that the average net forces on the binary and on the single effective point mass are commensurate when embedded in the same CE flow. In general, the flow morphology becomes less laminar when the binary is present. Predictably, the binary case shows modulations on the net force induced by its orbital motion. The periodic flow disturbances induced by the binary, which propagate at large distances, modify the density distribution around the bow shock region. These gas disturbances are clearly seen in the flow visualizations of Figures 6 and 4. The single point mass case, on the other hand, generates a stable bow shock structure once steady state is achieved (M. MacLeod et al. 2017). The density disturbances of the flow at large radii as induced by the binary lead to a slight decrease in the gas drag force when compared to the single point mass case (see differences between dotted lines and solid lines in the left panel of Figure 5).

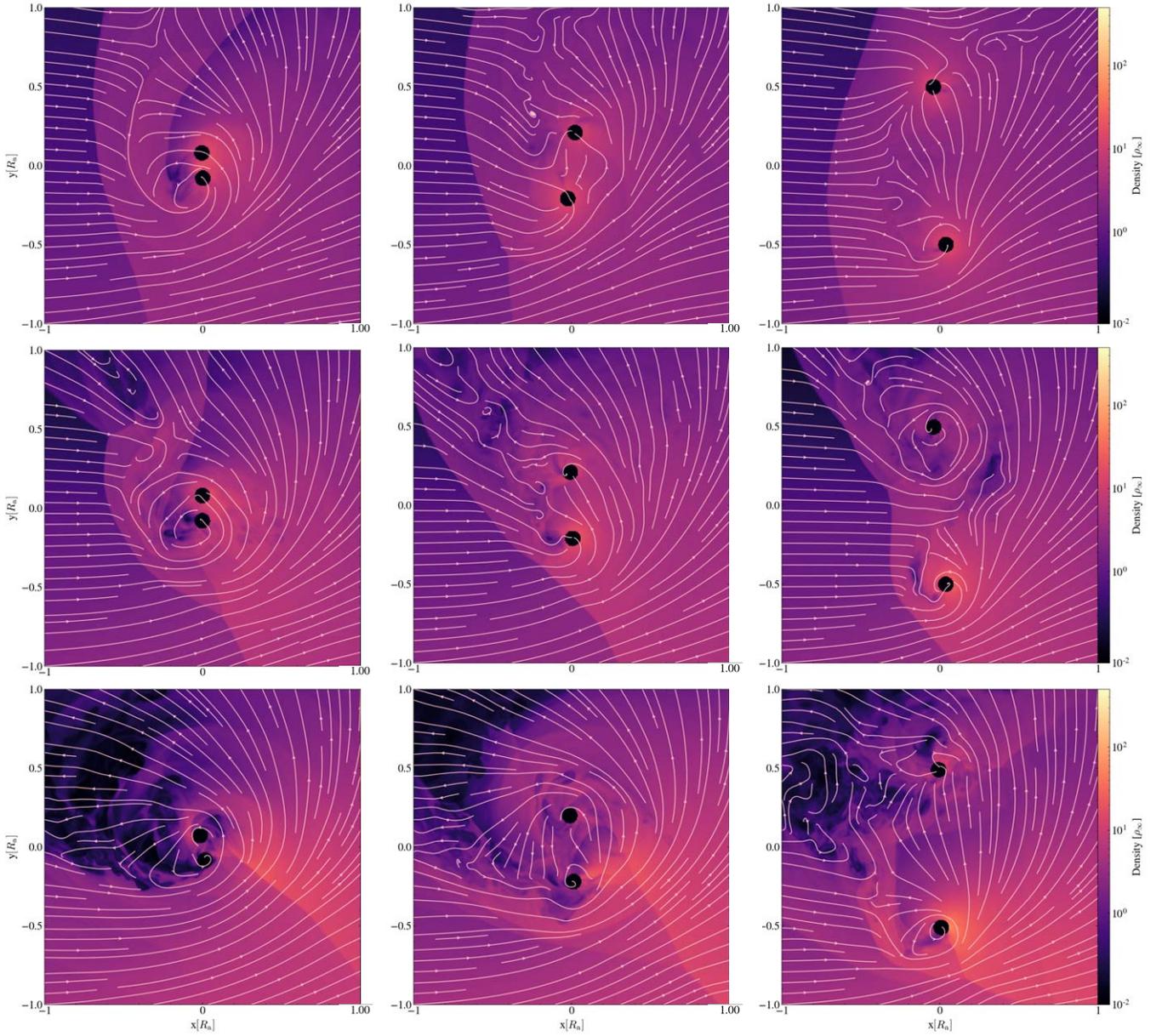

**Figure 4.** Flow morphology around the binary for the suite of simulations used in this study. Colormaps show the density of the flow in the orbital plane. The three different columns from left to right show the change in semimajor axis: $a_{b,0} = 0.16$ (left), 0.42 (middle), and 1.0 (right). The three different rows show the change in the flow driven by changes in $\varepsilon_\rho$: 0.5 (top), 1.0 (middle), and 2.0 (bottom).

Our numerical approach substitutes the binary stellar members with sinks on the grid, which absorb the convergent flow. As such, we investigate the properties of gas accreting through this inner boundary and note that the accretion rate is reliant on the size of the sink (M. MacLeod et al. 2017; A. Antoni et al. 2019). The right panel of Figure 5 shows the total mass accretion rate for both the binary (dark blue) and the effective single point mass (light blue). The accretion rate of the binary is the sum of the individual accretion rates for each of the stellar members. The mass accretion rate for the single point mass is consistent with the values derived by M. MacLeod et al. (2017) and is higher than the accretion rate of the binary. This is because, at the chosen binary separation ($a_{b,0} = 0.42 R_a$), the effective cross section of the single point mass is larger by about 20%. In the limit where $a_{b,0} \gtrsim R_a$, the binary system evolves according to the prescription for two independent objects (each of mass $\frac{1}{2}m$). In this limit, we would expect the accretion rate of the system to be $\propto (m_1^2 + m_2^2) = m^2/2$. Having demonstrated that the net drag force on the center of mass of the binary closely tracks the drag behavior of the single-object case, in what follows we investigate the impact that $\varepsilon_\rho$ has on the evolution of the binary.

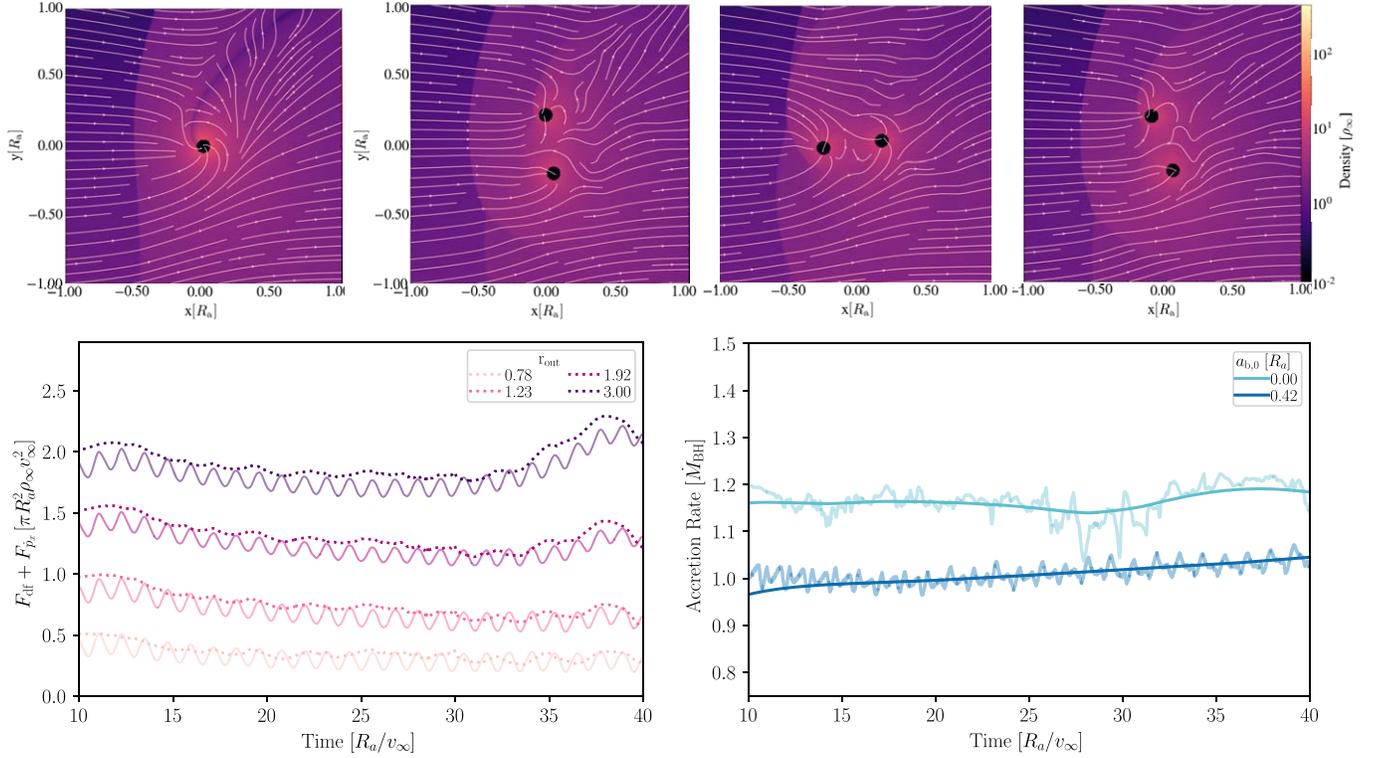

**Figure 5.** Net drag force (with contributions from dynamical friction and accretion of linear momentum) projected along the direction of the center-of-mass velocity vector and mass accretion rate as a function of simulation time for a binary embedded in a CE flow. The simulation parameters are $a_{b,0} = 0.42 R_a$, $\varepsilon_\rho = 0.3$, and $\mathcal{M} \approx 1.0$. The period of the binary is $\sim 1.2 R_a/v_\infty$. The effect on the binary is compared with that on a single point mass ($a_{b,0} = 0$). Top left panel: flow morphology for a single point mass. Top rightmost three panels: flow morphology for the binary case at three different times increasing from left to right. Density slices are taken for $z = 0$ plane surrounding the sinks. Bottom left panel: net drag force on the center of mass of the binary. The solid line shows the net force on the binary, while the dotted line shows the net force on the single mass particle. The different colors denote the distance from the center of mass to the integration radius. Bottom right panel: mass accretion rate for the single point mass and total accretion rate for the two sinks in the case of the binary. Lighter lines show the rate at each time step, and darker lines show the median values. The variations in the mass accretion rates are mainly driven by the periodic motion of the binary, which in turn modifies the density structure of the flow.

### 5.2. The Role of $\varepsilon_\rho$

As density gradients steepen, we thus expect the accretion rate to decrease and the net drag force to increase (M. MacLeod et al. 2017). Figure 6 illustrates the effects of varying $\varepsilon_\rho$ for a binary with $a_{b,0} = 0.42 R_a$ and $q = 0.15$, corresponding to $\mathcal{M} \approx 1.5, 2.0, 3.0$ for $\varepsilon_\rho = 0.5, 1,$ and 2, respectively. Three values of $\varepsilon_\rho$, which are representative of CE flows (Figure 2), are displayed: 0.5, 1, and 2. Our simulations are consistent with the flow properties studied by M. MacLeod et al. (2017). For a relatively shallow density gradient ($\varepsilon_\rho = 0.5$), the symmetry of the bow shock is just about maintained. As the density gradient steepens, the material focused onto the wake comes from material originating deeper within the stellar envelope, altering the shock morphology.

The bottom left panel in Figure 6 shows the net force on the center of mass of the binary. Here, we add the momentum transfer force on each of the sinks and the gaseous drag force integrated up to a distance $3.0 R_a$. As in M. MacLeod et al. (2017), the total drag force is observed to augment with increasing $\varepsilon_\rho$. This is because the drag force depends not only on the value of the density ($\rho_\infty$) at the position of the center of mass of the binary (A. Antoni et al. 2019) but also the mass redistribution within the accretion radius. As $\varepsilon_\rho$ increases, dense material from deeper within the stellar envelope is focused into the wake of the embedded binary, causing it to provide an increasing contribution to the net dynamical friction.

We now consider the forces impacting the motion of the binary about the center of mass. In the bottom right panel of Figure 6, we show the change in the binary separation, $\Delta a_b(t) = a_b(t) - a_{b,0}$, as a function of simulation time. The periodic deviations around the mean value are caused by the changing direction of the barycentric velocities over each single orbit. As in A. Antoni et al. (2019), gaseous drag forces can cause a tightening of the orbit when the density gradient is shallow ($\varepsilon_\rho = 0.5$). This is because the convergence of the gravitationally focused gas leads to a central density enhancement (when compared to $\rho_\infty$) in and around the binary orbit. As the density gradient steepens, material originating from deeper within the stellar envelope notably alters the density distribution of the bow shock region. As a result, the scaling from the gaseous drag forces evolves from a tightening to a softening of the binary at a critical $\varepsilon_\rho$, which, as we discuss in Section 5.5, also depends sensitively on $q$.

### 5.3. Key Timescales

A binary embedded in a CE will evolve as a result of drag and accretion. Forces on the center of mass lessen the kinetic energy content of the system, $E_{\text{trans}}$, over a timescale $\tau_{\text{stop}}$. The orbit-averaged decrease in kinetic energy content, $\langle \dot{E}_{\text{trans}} \rangle$, causes the sinking of the center of mass of the binary over a

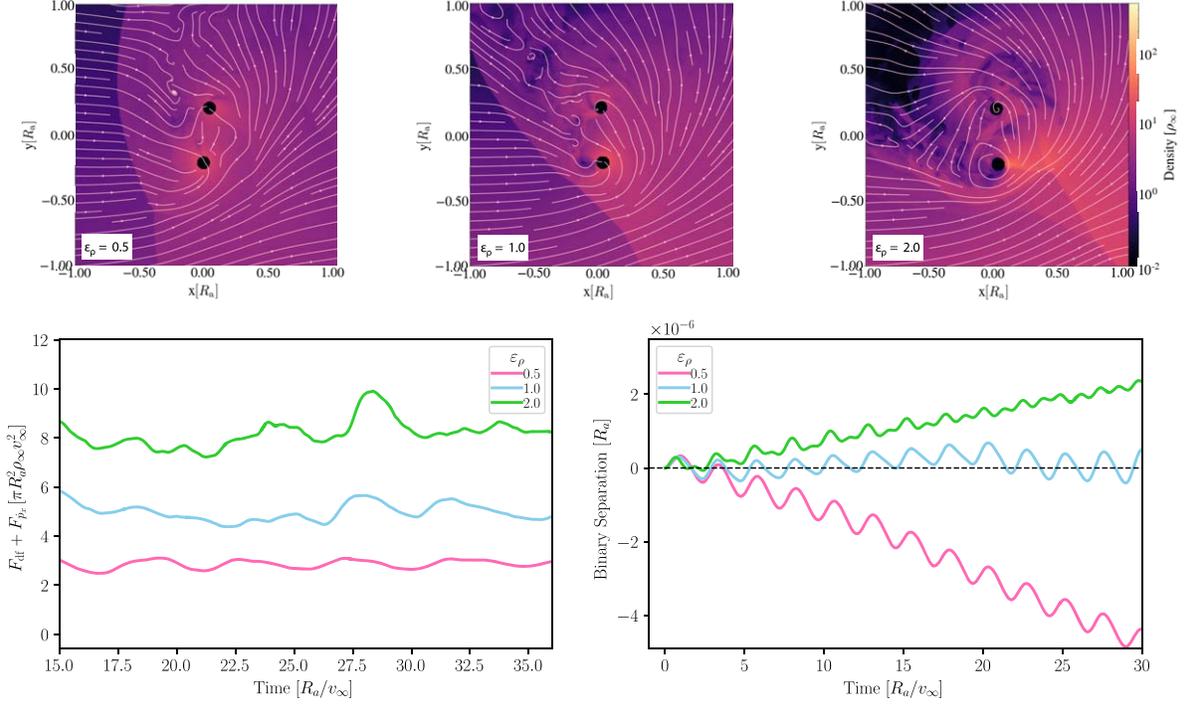

**Figure 6.** Diagnostics of CE flow properties. Top panels: flow morphology for a binary with $a_{b,0} = 0.42$ and $q = 0.15$ embedded in a CE flow with varying stellar density gradients. Density slices through the $z = 0$ plane surrounding the binary are displayed. From left to right, $\varepsilon_\rho = 0.5$, 1.0, and 2.0. Bottom left panel: the net drag force on the center of mass of the binary projected onto the velocity of the center-of-mass vector, including contributions from dynamical friction, $F_{df}$, and accretion of linear momentum, $F_{\dot{p}}$. The drag forces are integrated over a spherical domain up to a radius $3.0\,R_a$ from the center of mass of the binary. Bottom right panel: the change in the binary separation, $\Delta a_b(t) = a_b(t) - a_{b,0}$, as function of simulation time in units of $R_a$. For the case of a steep $\varepsilon_\rho$, the binary separation increases over time, while for a shallower $\varepsilon_\rho$, the separation decreases, as also found in the case of $\varepsilon_\rho = 0$ (A. Antoni et al. 2019).

timescale

$$\langle \tau_{\rm stop} \rangle \equiv \frac{E_{\rm trans}}{\langle \dot{E}_{\rm trans} \rangle}. \quad (19)$$

Forces on the binary orbit modify the separation over a timescale $\tau_{\rm insp}$. Both drag and momentum transport cause a net drag on the orbital motion. The orbit-averaged inspiraling timescale can thus be written as

$$\langle \tau_{\rm insp} \rangle \equiv \frac{-a_{b,0}}{\langle \dot{a}_b \rangle}, \quad (20)$$

where $\langle \dot{a}_b \rangle$ is calculated by taking the median over an integer number of orbits during the simulation. The sign of the timescale was chosen to be compatible with A. Antoni et al. (2019). For a binary that tightens, $\tau_{\rm insp} > 0$, while for softening binaries, $\tau_{\rm insp} < 0$. The magnitude of $\tau_{\rm insp}$ is thus a measure for how quickly the tightening or softening of the binary orbit takes place. For nonevolving binaries, $\tau_{\rm insp} \to \infty$, or equivalently, $\tau_{\rm insp}^{-1} \to 0$. In this section, we explore these timescales and their hierarchy, which will determine how a triple system evolves.

The sinking timescale of the center of mass inside the envelope, $\tau_{\rm stop}$, will determine how swiftly the binary reaches its tidal radius as it approaches the primary's He core. We find that $\tau_{\rm stop}$ is primarily determined by the properties of the flow ($q$ and $\varepsilon_\rho$). Similar to the $\varepsilon_\rho = 0$ case (A. Antoni et al. 2019), we find $\tau_{\rm stop} < \tau_{\rm insp}$ for $\varepsilon_\rho = 0.5$ and a systematic decrease in $\tau_{\rm stop}$ with increasing $\varepsilon_\rho$, as found in M. MacLeod et al. (2017) for the $a_{b,0} = 0$ case. The reader is referred to Figure 8 and

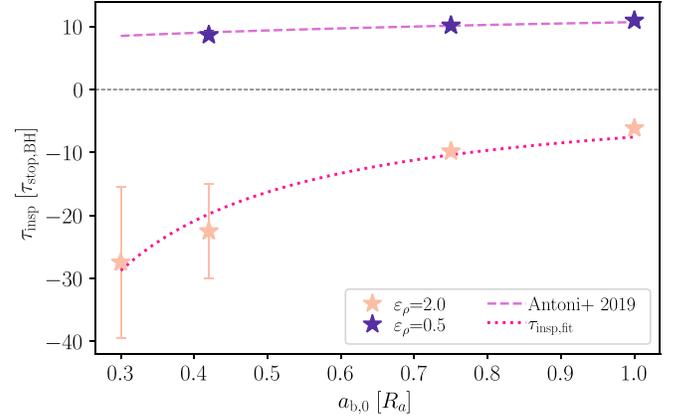

**Figure 7.** Orbit-averaged inspiraling timescale as a function of $a_{b,0}$, given by Equation (20). Values are normalized to the analytical BH stopping time given in Equation (7). For $\varepsilon_\rho = 0.5$, $\tau_{\rm insp} > 0$, while for $\varepsilon_\rho = 2$, $\tau_{\rm insp} < 0$. A positive timescale corresponds to an orbital tightening, while a negative timescale corresponds to an orbital widening. An increase in the absolute value of $\tau_{\rm insp}$ is associated with a decrease in the rate of orbital tightening or widening. A least-squares fit to a power law of the form $\propto (a_{b,0}/R_a)^B$ is performed. The shallow density gradient models ($\varepsilon_\rho = 0.5$) are compared with the trend $(a_{b,0}/R_a)^{0.19}$ found by A. Antoni et al. (2019). For the case where $\varepsilon_\rho = 2.0$, $\tau_{\rm insp} \propto (a_{b,0}/R_a)^{-1.1}$. This implies that binaries embedded in steep stellar density gradients will soften at a rate that progressively increases as $a$ expands.

Section 5.4 for a more in-depth presentation and discussion of our results, respectively.

Figure 7 depicts the dependence of $\tau_{\rm insp}$ as a function of $a_{b,0}$ for both shallow and steep stellar density gradient models. We calculate the $a_{b,0}$ dependence of $\tau_{\rm insp}$ by performing a least-squares fit to a power law of the form $\propto (a_{b,0}/R_a)^B$. The shallow density gradient models ($\varepsilon_\rho = 0.5$) are compared with the scaling of $\tau_{\rm insp}$ with $a_{\rm b,0}$ of $B \approx 0.19$ derived by A. Antoni et al. (2019) for $\varepsilon_\rho = 0$. The orbital evolution with separation thus follows the scaling found for a constant-density medium, although the orbit-averaged torques are weaker. We find that, for shallow density gradients, the inspiral timescale ($\tau_{\rm insp} > 0$) decreases with $a_b$ such that the system will merge if given enough time.

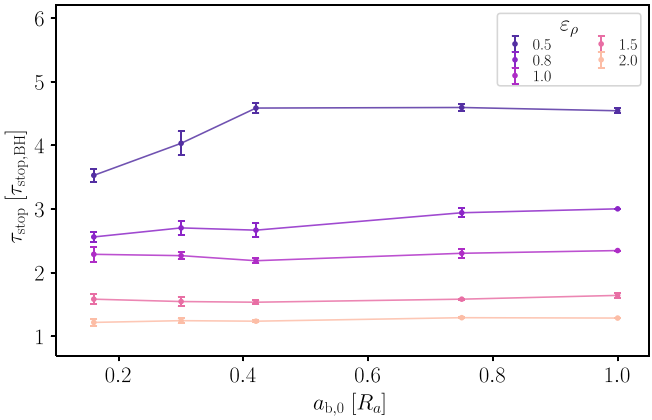

**Figure 8.** Stopping timescale for a binary with $q = 0.15$, embedded in a CE flow of varying physical conditions. Values are normalized by $\tau_{\rm stop,\,BH}$ as in Equation (7). The colors represent the different values of the density gradient parameter: $\varepsilon_\rho$. The trend shows that the stopping timescale of the center of mass of the binary decreases as $\varepsilon_\rho$ increases, so the center of mass of the binary will spiral in faster into the CE. We find $\tau_{\rm stop}$ to remain fairly constant regardless of the initial separation of the binary.

In contrast, for steep density gradients ($\varepsilon_\rho = 2.0$), the separation of the binary is found to increase with time ($\tau_{\rm insp} < 0$) at all separations. As such, we expect the binary will widen as it sinks into the surrounding envelope. As the binary separation increases, the orbital speed of each object decreases, augmenting the total outward torque. In other words, as the binary softens, the magnitude of $\tau_{\rm insp}$ becomes progressively less negative, which implies that binary expansion is gradually accelerated. For $\varepsilon_\rho = 2.0$, we find $B \approx -1.1$ (Figure 7).

### 5.4. The Sinking of the Center of Mass During CE

The stopping timescale, $\tau_{\rm stop}$, is a measure of the loss of kinetic energy of the center of mass of the binary (Equation (7)). $\tau_{\rm stop}$ in the CE context is a measure of how quickly the center of mass of the binary will spiral in toward the core of the evolving primary star. For our simulated binaries, we measure orbit-averaged stopping timescales. These are plotted in Figure 8. The values are normalized to the analytical solution for a single point mass found in Equation (7). We see that the general trend is that the stopping timescale becomes shorter as $\varepsilon_\rho$ increases. We can also see that the stopping timescale remains pretty much constant with respect to the initial separation $a_{\rm b,0}$. The dissipation of the kinetic energy therefore depends primarily on the density gradient of the stellar profile. Our findings are in close agreement with the results of M. MacLeod et al. (2017) for single point masses.

### 5.5. The Hardening or Softening of Binaries During CE

In what follows, we further explore the conditions for which $\tau_{\rm insp}$ reverts sign in the ($\varepsilon_\rho$, $a_{\rm b,0}$) plane. In the simulation results presented in Figure 7, we fix $q = 0.15$ and vary $a_{\rm b,0}$ and $\varepsilon_\rho$. Shown is the orbit-averaged frequency $\tau_{\rm insp}^{-1}$, which has been selected here in lieu of $\tau_{\rm insp}$ in order to more effectively depict binaries that show no appreciable change in their separation over the duration of the simulations. The values in the colorbar are normalized to the analytical BH stopping time given in Equation (7). The pink color palette corresponds to binaries that tighten ($\tau_{\rm insp}^{-1} > 0$), while the blue color palette shows binaries that soften ($\tau_{\rm insp}^{-1} < 0$). Small absolute frequencies, $|\tau_{\rm insp}^{-1}|$, correspond to binaries experiencing a slower rates of change for both tightening and softening. The symbols in white depict binaries that experience no appreciable change (within simulation errors) in their orbital separation ($\tau_{\rm insp}^{-1} \to 0$) and thus demarcate transitional flow conditions in the ($\varepsilon_\rho$, $a_{\rm b,0}$) plane.

Figure 7 shows that, for $\varepsilon_\rho < 1$, binaries harden, while softening takes place when $\varepsilon_\rho > 1.0$. We note that this transition also depends on $a_{\rm b,0}$, albeit weakly compared to $\varepsilon_\rho$. In the presence of a steep density gradient, the two members of the binary spiral out as the flow transfers angular momentum to the system. For a fixed $\varepsilon_\rho$, the long-term evolution of a binary system can be envisioned by moving along vertically in the ($\varepsilon_\rho$, $a_{\rm b,0}$) plane. As observed in Figure 7 for the $\varepsilon_\rho = 0.5$ case and discussed by A. Antoni et al. (2019) for the $\varepsilon_\rho = 0$ case, the decrease in $\tau_{\rm insp}$ with $a_b$ indicates that gaseous drag forces in shallow density gradients can drive binaries to coalescence or to the critical separation at which gravitational-wave radiation controls their subsequent orbital evolution. In contrast, for steeper density gradients ($\varepsilon_\rho > 1$), we find that, as the two binary members recede, the rate of softening increases, hastening the potential for tidal breakup. Given that, for most CE flows, $\varepsilon_\rho \gtrsim 1$ (R. W. Everson et al. 2020), we thus conclude that binaries spiraling within CEs become less bound and are more vulnerable to be torn apart by tidal interactions as they fall closer to the primary's core.

### 5.6. The Assembly of PSR J0337+1715

The evolution of triples is likely to be important for the formation of compact and exotic binaries. The evolution of a triple system with a mass-transferring outer star has been invoked to explain the triple system PSR J0337+1715 (T. M. Tauris & E. P. J. van den Heuvel 2014; E. Sabach & N. Soker 2015), which hosts a millisecond pulsar and two white dwarfs. This system is used in Section 2 to frame the typical flow conditions encountered by embedded CE binaries, and it has been argued that they are indeed representative of the bulk population of CE interactions (R. W. Everson et al. 2020).

To contextualize our numerical results (Figure 9) in the context of PSR J0337+1715, we find that $\varepsilon_\rho$ in the envelope of the outer star varies between roughly 0.98 and 1.63 as a function of the distance from the center of mass of the evolving massive stellar progenitor. The reader is referred to Figure 3 for further details. This illustrates that the CE properties encounter by the embedded binary progenitor system resides on the right side of the parameter space explored in Figure 9. This in turn suggests that the embedded binary will soften and will become more vulnerable to tidal disruption. This provides credence to the idea by E. Sabach & N. Soker (2015) that tidal breakup of the binary can alter the hierarchical layout of this system. Based on the results of this work, we predict that binary splitting is inevitable and that the softening of the binary as it spirals near the tidal radius will only aid breakup.

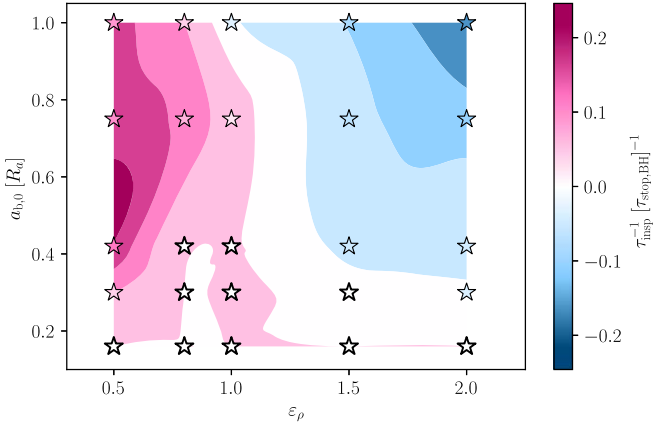

**Figure 9.** The orbit-averaged frequency $\tau_{\rm insp}^{-1}$ in the $(a_{b,0}, \varepsilon_\rho)$ plane for a binary with $q = 0.15$. $\tau_{\rm insp}^{-1}$ provides an estimate of the ratio between the change in binary separation $\Delta a_b$ averaged over an orbital period and the initial separation $a_{b,0}$. The orbit-averaged frequency $\tau_{\rm insp}^{-1}$ is selected here to better illustrate cases that, over the duration of the simulation, show no appreciable change in their separation. For such cases, $\tau_{\rm insp}^{-1} \approx 0$. Values in the colorbar are given in units of the analytical BH stopping time given in Equation (7). The pink-colored values show binaries that are tightened by the interaction with the CE flow ($\tau_{\rm insp}^{-1} > 0$), while blue colors indicate binaries that become less gravitationally bound ($\tau_{\rm insp}^{-1} < 0$). The darker colors mark binaries that experience a larger change in $\Delta a_b$ over an orbital period. The white symbols show cases where $\Delta a_b \approx 0$ within the simulation errors, and thus mark the transition between binary hardening and binary softening.

## 6. Summary and Concluding Remarks

Field stars are commonly formed in pairs, and a fraction of these binaries are part of triple systems. Some binary members are expected to endure mass exchanges and can, mainly during a phase of unstable mass transfer, spiral into the interior layers of the evolving star under the influence of drag from the surrounding envelope material. The associated drag forces deposit orbital kinetic energy into the envelope and are expected to influence the orbital configuration of the infalling binary in ways that remain unknown.

In this paper, we present a set of simulations that model a binary system as it becomes embedded into the CE of their tertiary. We approach this problem by running local hydrodynamic simulation in a wind tunnel setup and focus on a range of parameters informed by observations of hierarchical triple systems. We study the forces on the center of mass of the binary with contributions from dynamical friction and accretion of linear momentum. These numerical experiments have the advantage of being sufficiently controlled that we can gain full confidence in the numerical implementation and intuition for the effects of key variables. Our calculations have indicated that the structure of a CE has dramatic importance in shaping the properties of the embedded binary. Some key findings of this work are as follows.

1. The forces on the center of mass of the binary are found to be comparable in magnitude and direction to those experienced by an effective star of mass $m$ at the center of mass of the inner binary. We thus infer that the binary will sink into the envelope at a rate comparable to that previously found for single point masses (M. MacLeod et al. 2017).
2. The net accretion rate of the two bodies sees a reduction compared to that of a single point mass, which can be readily explained by the difference in accretion cross sections.
3. The orbital evolution of binaries embedded in shallow density gradients ($\varepsilon_\rho < 0.5$) is found to follow the scaling with separation derived by A. Antoni et al. (2019) for a constant-density medium ($\varepsilon_\rho \approx 0$). Such density gradients are not commonly found in the interiors of evolving stars.
4. In contrast to the uniform density case, the binary orbital separation is found to increase with time for $\varepsilon_\rho > 1$. This causes binaries in CEs to soften and prevents them from subsequently merging. For these customary CE properties, the binary softens and the magnitude of the timescale decreases as the separation increases, which subsequently accelerates the softening. We conclude that, for realistic CE conditions, embedded binaries will soften, thus aiding disruption and ejection.

This study addresses a critical stage in the evolution of hierarchical triple systems: common envelope episodes. Given that multiple-star systems are ubiquitous, we urgently need a more detailed understanding of the binary-star interaction channels to be able to more effectively decipher how the majority of triple systems evolve. This work tackles key aspects of the basic physics of binary–CE interactions. Triple systems might be interesting sources for LIGO progenitors (e.g., A. Vigna-Gómez et al. 2021) and might be central to understanding type Ia supernova explosions (e.g., A. S. Rajamuthukumar et al. 2023), and without a detailed understanding of CE, it might not be feasible to precisely determine the distribution of their properties.

## Acknowledgments

We thank the referee for the careful and insightful review of our manuscript. We acknowledge use of the lux supercomputer at UC Santa Cruz, funded by NSF MRI grant AST 1828315. The 3D hydrodynamics software used in this work was developed in part by the DOE NNSA- and DOE Office of Science-supported Flash Center for Computational Science at the University of Chicago and the University of Rochester. E. R.-R. acknowledges support by the Heising-Simons Foundation and the NSF (AST-2307710, AST-2206243, AST-1911206, and AST-1852393). R.Y. is grateful for support from a Doctoral Fellowship from the University of California institute for Mexico and the United States (UCMEXUS) and the Consejo Nacional de Ciencia y Tecnología (CONACyT), a Texas Advanced Computing Center (TACC) Frontera Computational Science Fellowship, and a NASA FINESST award (21-ASTRO21-0068). A.M-B. is supported by NASA through the NASA Hubble Fellowship grant HST-HF2-51487.001-A awarded by the Space Telescope Science Institute, which is operated by the Association of Universities for Research in Astronomy, Inc., for NASA, under contract NAS5-26555. R. W.E. acknowledges the support of the University of California President's Dissertation-Year Fellowship, the Heising-Simons Foundation, the Vera Rubin Presidential Chair for Diversity at UCSC, and the National Science Foundation Graduate Research Fellowship Program under grant No. 1339067. A. A. gratefully acknowledges support from the University of California, Dissertation-Year Fellowship, the Maria Cranor Fellowship at U.C. Berkeley, the National Science Foundation Graduate Research Fellowship under grant No. DGE 1752814, and the Gordon and Betty Moore Foundation through grant GBMF5076.

*Software*: FLASH 4.3 (B. Fryxell et al. 2000), MESA (J. R. Buchler & W. R. Yueh 1976; G. M. Fuller et al. 1985; C. A. Iglesias & F. J. Rogers 1993; T. Oda et al. 1994; D. Saumon et al. 1995; C. A. Iglesias & F. J. Rogers 1996; N. Itoh et al. 1996; K. Langanke & G. Martínez-Pinedo 2000; F. X. Timmes & F. D. Swesty 2000; F. J. Rogers & A. Nayfonov 2002; J. W. Ferguson et al. 2005; A. I. Chugunov et al. 2007; S. Cassisi et al. 2007; A. Y. Potekhin & G. Chabrier 2010; R. H. Cyburt et al. 2010; A. W. Irwin 2012; B. Paxton et al. 2011, 2013, 2015, 2018, 2019; A. S. Jermyn et al. 2023), MESA SDK (R. Townsend 2021), yt (M. J. Turk et al. 2011).

**ORCID iDs**

Alejandra Rosselli-Calderon https://orcid.org/0000-0002-9537-1933
Ricardo Yarza https://orcid.org/0000-0003-0381-1039
Ariadna Murguia-Berthier https://orcid.org/0000-0003-2333-6116
Valeriia Rohoza https://orcid.org/0009-0000-4703-9808
Rosa Wallace Everson https://orcid.org/0000-0001-5256-3620
Andrea Antoni https://orcid.org/0000-0003-3062-4773
Morgan MacLeod https://orcid.org/0000-0002-1417-8024
Enrico Ramirez-Ruiz https://orcid.org/0000-0003-2558-3102

# Appendix
## Application of our Results to a Broad Range of Embedded Binaries

The characteristic Mach number of the flow $\mathcal{M}_\infty$ is determined by $\varepsilon_\rho$ and $q$ (Equation (12)). For most of the analysis in the study, we have selected $q = 0.15$, inspired by the post-CE formation paradigm of PSR J0337+1715. Here, we expand our analysis to include two different values of $q$. Figure 10 shows the results of our study. We color the symbols depending on whether the final orbital separation is greater (blue) or smaller (pink) than $a_{b,0}$. For binaries in which the separation increases (decreases), we label them as breaking (merging). The general trend observed here is similar for both $q = 0.10$ and $q = 0.15$. We observe a clear transition at around $\varepsilon_\rho = 1.0$, with steeper (shallower) density gradients causing binaries to soften (harden). The relation between $\mathcal{M}_\infty$, $\varepsilon_\rho$, and $q$ (Equation (12)) is clearly displayed by the gray lines in Figure 10.

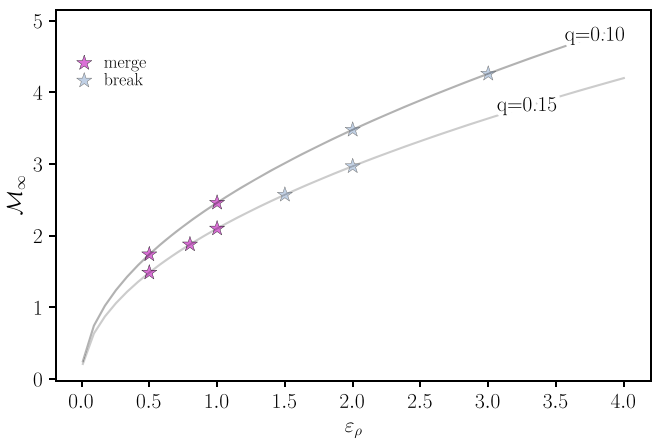

**Figure 10.** The evolution of binaries in the ($\varepsilon_\rho$, $\mathcal{M}_\infty$) plane using numerical results from the simulations presented in this study. The relation between $\mathcal{M}_\infty$, $\varepsilon_\rho$, and $q$ (Equation (12)) is shown for $q = 0.10$ and $q = 0.15$ (gray lines). The blue (pink) symbols show cases where binaries harden (soften) as they interact with the CE flow.